\documentclass[aps,article,showpacs,showkeys]{revtex4}
\usepackage{amsfonts}
\usepackage{amsmath}
\usepackage{amssymb}
\usepackage{graphicx}

\renewcommand{\,}{{}}
\renewcommand{\a}{\alpha}
\renewcommand{\b}{\beta}

\newcommand{\D}{\Delta}
\renewcommand{\o}{\omega}

\newcommand{\g}{\gamma}
\newcommand{\G}{\Gamma}
\newcommand{\m}{\mu}
\newcommand{\n}{\nu}
\newcommand{\e}{\epsilon}
\newcommand{\x}{\xi}

\renewcommand{\mathbf}{\pmb}

\renewcommand{\L}{{\cal L}}
\newcommand{\M}{\mathcal{M}}
\newcommand{\mt}{{g}}

\renewcommand{\)}{\right)}
\renewcommand{\[}{\left[}
\renewcommand{\]}{\right]}
\newcommand{\aeq}{\begin{equation}}
\newcommand{\ceq}{\end{equation}}
\newcommand{\aec}{\begin{eqnarray}}
\newcommand{\cec}{\end{eqnarray}}
\newcommand{\ptm}{{\mathbf P}^{\left(\frac{3}{2}\right)}}
\newcommand{\pum}{{\mathbf P}^{\left(\frac{1}{2}\right)}}
\makeindex
\begin{document}

\title{Compton scattering off elementary spin $\frac{3}{2}$ particles}
\author{E. G. Delgado-Acosta and M. Napsuciale }
\affiliation{Departamento de F\'{\i}sica, Divisi\'{o}n de Ciencias e Ingenier\'{\i}as,
Universidad de Guanajuato, Campus Le\'{o}n, Lomas del Bosque 103,
Fraccionamiento Lomas del Campestre, 37150, Le\'{o}n, Guanajuato, M\'{e}xico.}

\begin{abstract}
We calculate Compton scattering off an elementary spin $\frac{3}{2}$
particle in a recently proposed framework for the description of 
high spin fields based on the projection onto eigen-subspaces of the Casimir 
operators of the Poincar\'{e} group. We also calculate this process in the conventional 
Rarita-Schwinger formalism. Both formalisms yield 
the correct Thomson limit but the predictions for the angular distribution and total 
cross section differ beyond this point. We point out that the average squared amplitudes 
in the forward direction for Compton scattering off targets with spin $s=0,\frac{1}{2},1$ are 
energy-independent and have the common value $4e^{4}$. As a consequence, in the rest 
frame of the particle the differential cross section for Compton scattering in the 
forward direction is energy independent and coincides with the classical squared radius. 
We show that these properties are also satisfied by a spin $3/2$ target in the Poincar\'{e} 
projector formalism but not by the Rarita-Schwinger spin $3/2$ particle.
\end{abstract}


\maketitle

\section{Introduction} 
A long standing problem in particle physics is the proper description 
of high spin fields. The widely used Rarita-Schwinger (RS) formalism \cite{RS} 
was shown to be inconsistent for interacting particles long ago \cite{JS}, and 
lead to superluminical propagation of spin 3/2  waves in the presence of an 
external electromagnetic field  \cite{VZ}. Similar and related problems have been found 
in the presence of other interactions \cite{todos}. 

Recently, a new formalism for the description of high spin fields was put forward 
\cite{NKR}(NKR formalism in the following), based on the projection onto eigensubspaces of the 
Casimir operators of the Poincar\'{e} group. In that work, it is shown that, under 
minimal coupling, the (parity conserving) electromagnetic structure of a spin 3/2 
particle transforming in the 
$(1/2,1/2)\otimes [(1/2,0)\oplus (0,1/2)]$ representation of the Homogeneous Lorentz 
Group (HLG) depend on two free parameters denoted $g$ and $f$. The propagation of 
spin 3/2 waves was studied for the case $f=0$ and it is shown there that the value of 
the gyromagnetic factor $g$ is related to the causality of the propagation of 
spin 3/2 waves and causal propagation is obtained for $g=2$. This result relates the 
``natural" value of the gyromagnetic factor \cite{naturalg} to causality for spin $3/2$.

The case of spin 1 particles in the $\left(1/2,1/2\right)$ representation space of the HLG 
was addressed in \cite{NRDK}. In this case, 
the most general electromagnetic interaction of a spin 1 vector particle was also shown to 
depend on two parameters (denoted $g$ and $\xi$) which cannot be fixed from the Poincar\'{e} 
projection alone. These parameters determine the electromagnetic structure of the 
particle and were fixed imposing unitarity at high energies for Compton scattering. 
This procedure 
fixes the parameters to $g=2$ and $\xi=0$ predicting a gyromagnetic factor 
$g=2$, a related quadrupole electric moment $Q=-e(g-1)/m^{2}$ and vanishing odd-parity
couplings as a consequence of $\xi=0$  . The obtained couplings coincide with the ones 
predicted for the  $W$ boson in the Standard Model \cite{SM}. 

These results make worthy to study the analogous problems for spin 3/2 particles and 
this work is devoted to this purpose. The electromagnetic properties of spin $3/2$ particles 
has been addressed in a number of previous papers aiming to understand either the 
electromagnetic structure of hypothetical elementary particles or the 
electromagnetic properties of hadrons \cite{naturalg,let}. 

In this work we study the electromagnetic structure of a spin $3/2$ particle in the NKR 
formalism and calculate Compton scattering both in the NKR and RS formalisms. 
We compare the predictions of these formalisms for the angular distribution and total 
cross section and notice that the average squared amplitude for Compton scattering 
off spin $0,1/2$ and $1$ particles in the forward direction is energy independent. 
This property is satisfied by spin $3/2$ particles in the NKR formalism but not in the 
Rarita-Schwinger one. 
This paper is organized as follows: in the next section we revisit the electromagnetic 
structure of a spin 3/2 particle under $U(1)_{em}$ gauge principle in the NKR formalism, 
extract the corresponding Feynman rules and prove that Ward identities are satisfied.  
In section III we calculate the amplitude for Compton scattering, show that it is gauge 
invariant and work out the predictions for the differential and total cross sections. 
In section IV we calculate this process in the conventional Rarita-Schwinger formalism. 
We discuss our results in section V and give a summary in section VI.

\section{Electromagnetic interactions of spin 3/2 particles in the NKR formalism}

The NKR Lagrangian for spin 3/2 interacting particles with charge $-e$ has been 
discussed in \cite{NKR} and we refer the reader to this work for the details. 
The most general free Lagrangian for a spin 3/2 particle arising from the 
Poincar\'{e} projectors is
\aec
\label{lagvert}
\L_0(a,b)&=&(\partial^\m\overline{\psi}^\a)\G_{\a\b\m\n}\partial^\n\psi^\b
-m^2\overline{\psi}^\a\psi_\a+\frac{1}{a}(\partial^\m\overline{\psi}_\m)
(\partial^\a\psi_\a)+\frac{m^2}{b}(\overline{\psi}^\m\g_\m)(\g^\a\psi_\a).
\cec
Here, $a$, $b$ are free (``gauge") parameters and the corresponding (``gauge fixing") 
terms are associated to the constraints (see \cite{NKR} for a discussion on this point). 
The most general tensor compatible with Poincar\'{e} projection and Lorentz covariance is 
\begin{equation}  \label{gamageneral}
\Gamma_{\alpha\beta\mu\nu}=B_{\alpha\beta\mu\nu}-ig[M_{\mu\nu}]_{%
\alpha\beta}+\tilde{d} \gamma^5[M_{\mu\nu}]_{\alpha\beta} + \tilde{c}
\epsilon_{\alpha\beta\mu\nu}+if\gamma^5 \epsilon_{\alpha\beta\mu\nu}, 
\end{equation}
with
\begin{align} 
B_{\alpha\beta\mu\nu}=&\frac{1}{3} \left(-\gamma _{\beta }\gamma _{\nu
} g_{\alpha \mu }-2 g_{\beta \nu } g_{\alpha \mu }+\gamma _{\alpha }\gamma
_{\mu } g_{\beta \nu }-\gamma _{\alpha }\gamma _{\beta } g_{\mu \nu }+3
g_{\alpha \beta } g_{\mu \nu }\right), \\
[M_{\mu\nu}]_{\alpha\beta}=&\frac{1}{2} \sigma _{\mu \nu } g_{\alpha
\beta }+i \left(g_{\mu \alpha } g_{\nu \beta }-g_{\mu \beta } g_{\nu \alpha
}\right).
\end{align}
Here $M_{\mu\nu}$ are the generators of the $(1/2,1/2)\otimes[(1/2,0)\oplus(0,1/2)]$ representation 
of the H.L.G. and $\sigma_{\mu \nu }=\frac{i}{2}[\gamma_{\mu},\gamma_{\nu}]$. We included the 
odd-parity terms $\tilde{c},\tilde{d}$ for the sake of completeness. This tensor coincides with 
the one in Eq.(141) of \cite{NKR} when $f=0$ and $\tilde{c}=\tilde{d}=0$; it 
has been slightly rewritten for convenience in the calculations below.

The propagator is calculated as the inverse of the kinetic term. We obtain \cite{NKR}
\aeq
S(p,a,b)=\frac{\Delta(p,a,b)}{p^2-m^2+i\e},
\ceq
with:
\aec
\Delta(p,a,b)&=&-\ptm -\x \[\(bp^2+a(1-b)m^2\)\pum_{11}+a(3-b)m^2\pum_{22}-\sqrt{3}am^2(\pum_{12}+\pum_{21})\],\\
\x&=&\frac{b}{m^2}\frac{p^2-m^2}{(3-b)\(bp^2-a(1-b)m^2\)-3am^2}.
\cec
Here, $\ptm$ stands for the spin 3/2 projector and $\pum_{ij}$ are the spin 1/2 
projectors (for $i=j$) and ``switch" operators (for $i\neq j$) in the 
$(1/2,1/2)\otimes [(1/2,0)\oplus (0,1/2)]$ representation space of the HLG. 

\begin{figure}[ht]
\begin{minipage}{7cm}
\centering
\includegraphics[scale=0.25]{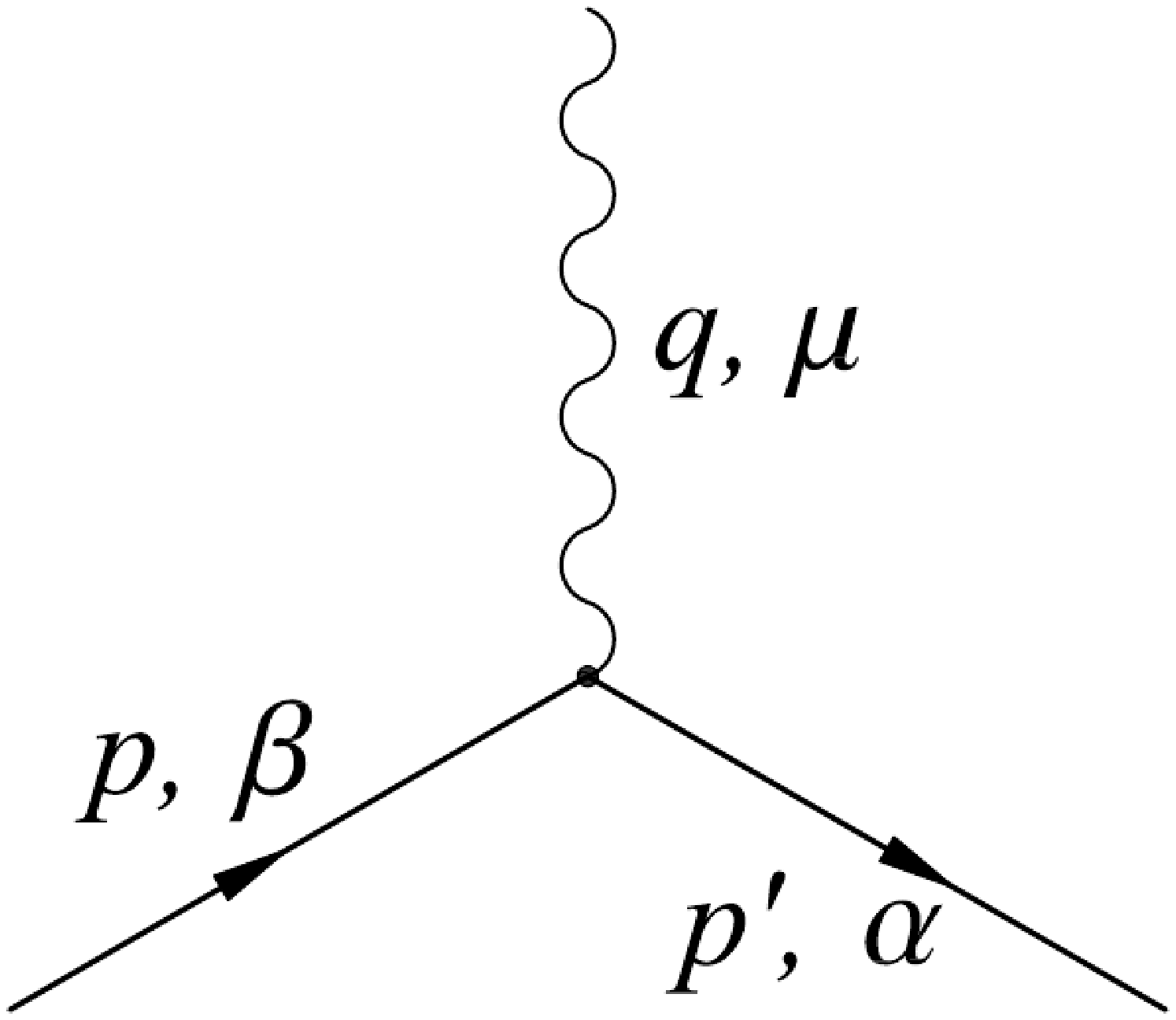}
\begin{equation*}
\label{reglad} 
e \[\G_{\a\b\n\m}p'^\n+\G_{\a\b\m\n}p^\n+\frac{1}{a}(\mt_{\m\a}p^\b+\mt_{\m\b}p'^\a)\]
\end{equation*}
\end{minipage}
\ \ \hfill
\begin{minipage}{7cm}
\centering
\includegraphics[scale=.25]{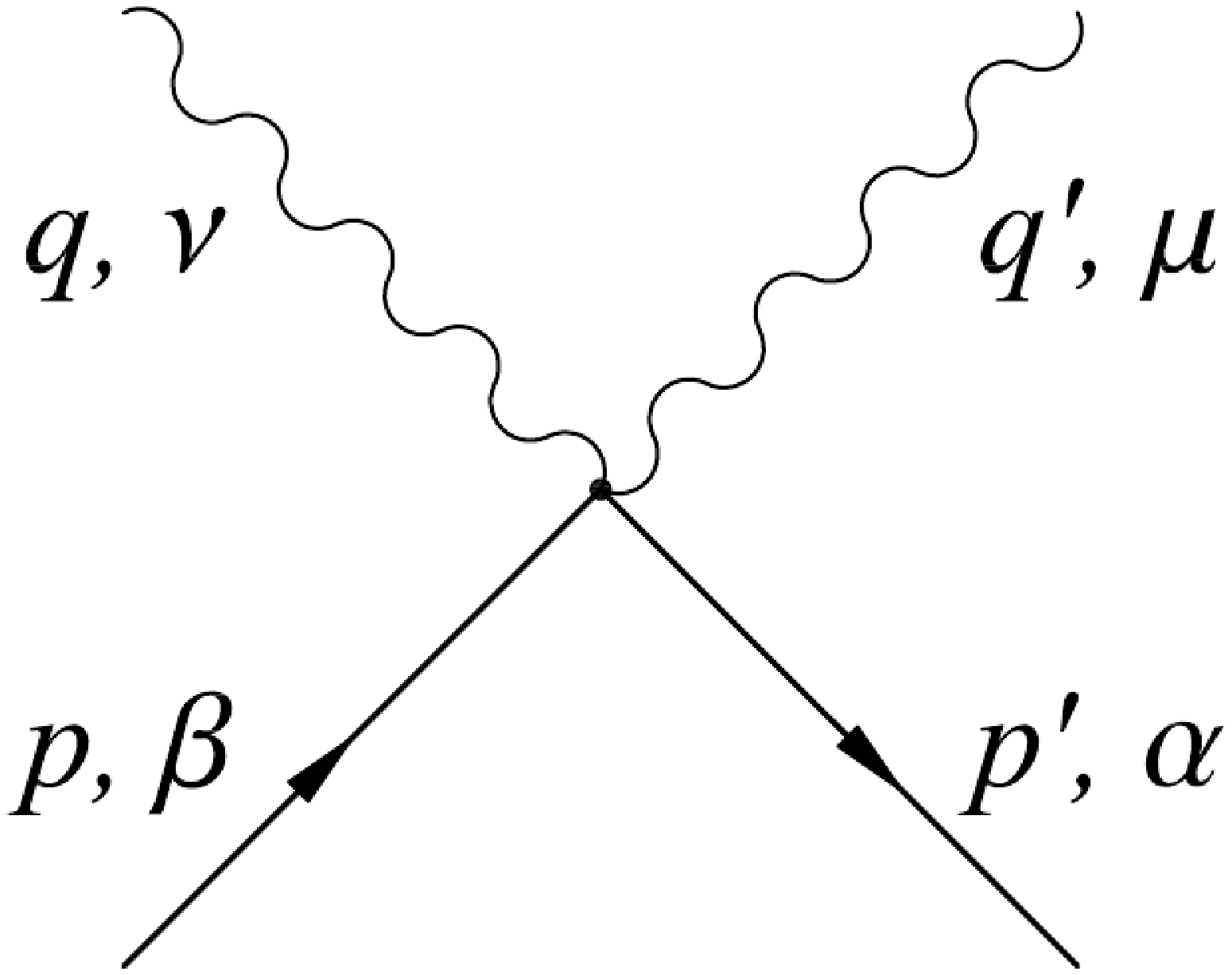}
\begin{equation*}
\label{reglac}
-\frac{1}{2}e^2\[\Gamma_{\alpha\beta\mu\nu}+\Gamma_{\alpha\beta\nu\mu}+\frac{1}{a}(\mt_{\a\m}\mt_{\b\n}+\mt_{\b\m}\mt_{\a\n})\]
\nonumber
\end{equation*}\end{minipage}\\
\vskip 1.0cm
\begin{minipage}{7cm}
\centering
\includegraphics[scale=0.25]{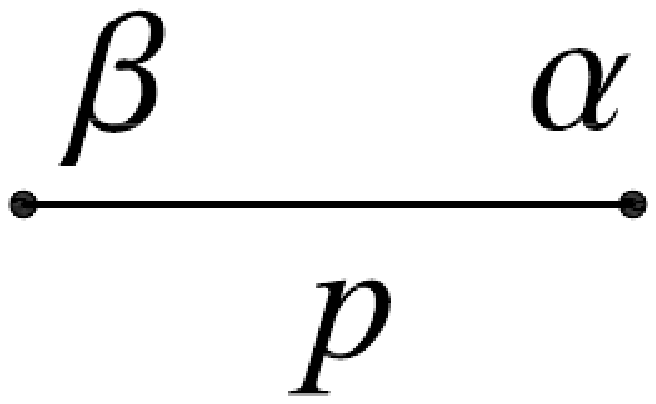}
\begin{equation*}\label{prop}
S_{\alpha\beta}(p,a,b)=\frac{\D_{\a\b}(p,a,b)}{p^2-m^2+i\e}
\end{equation*}
\end{minipage}
\caption{Feynman rules for arbitrary values of the ``gauge" parameters $a,b$.} 
\label{FR}
\end{figure}

Electromagnetic interactions are introduced in Eq. (\ref{lagvert}) using the $U(1)_{em}$ 
gauge principle which amounts to use the minimal coupling recipe 
$\partial^\a\rightarrow D^\a=\partial^\a-ieA^\a$. We obtain 
\aec
\L(a,b)=\L_0(a,b)-ej_\m(a)A^\m+e^2(\overline{\psi}^a\G_{\a\b\m\n}\psi^\b +
\frac{1}{a}\overline{\psi}_\m\psi_\n)A^\m A^\n,
\label{NKRint}
\cec
with
\aeq
j_\m(a)= i\overline{\psi}^\a\G_{\a\b\m\n}\partial^\n\psi^\b-i\partial^\n\overline{\psi}^\a \G_{\a\b\n\m}\psi^\beta+
\frac{1}{a}(\overline{\psi}_\m i\partial\cdot\psi-i\partial\cdot\overline{\psi}\psi_\m).
\ceq
In momentum space, using $\psi_\alpha=u_\alpha(p)e^{-ip\cdot x}$ the transition current reads
\begin{eqnarray}
j_\mu &=&\overline{u}^\alpha (p')\left[ \Gamma_{\alpha\beta\nu\mu}p'^\nu+\Gamma_{\alpha\beta\mu\nu}p^\nu +
\frac{1}{a}(g_{\mu\beta}p'_\alpha+g_{\mu\alpha}p_\beta)\right] u^\beta (p),\nonumber\\
&\equiv&\overline{u}^\alpha(p')\mathcal{V}(p',p,a)_{\alpha\beta\mu}u^\beta(p), 
\end{eqnarray}
where the electromagnetic vertex ${\cal V}(p',p,a)$ is defined by the latter relation. 
The Feynman rules derived from Eq. (\ref{NKRint}) are shown in Fig. (\ref{FR}).

A straightforward calculation shows that this vertex satisfy
\begin{align}
(p'-p)^\m{\cal V}&(p',p,a)_{\a\b\m}=\nonumber\\
&\left\lbrace K_{\a\b}(p')+\frac{1}{a}(p'_\a p'_\b)-m^2\mt_{\a\b}
+\frac{m^2}{b}\g_\a\g_\b  \right\rbrace-
\left\lbrace K_{\a\b}(p)+\frac{1}{a}(p_\b p_\a)-m^2\mt_{\a\b}
+\frac{m^2}{b}\g_\a\g_\b\right\rbrace,\label{vertnu}
\end{align}
where $K_{\a\b}(p)\equiv\G_{\a\b\m\n}p^\m p^\n$. In terms of the inverse propagator we get
\aeq
(p'-p)^\m{\cal V}(p',p,a)_{\a\b\m}=S^{-1}_{\alpha\beta}(p',a,b)-S^{-1}_{\alpha\beta}(p,a,b),
\ceq
i.e. the Ward-Takahashi identity is satisfied for any value of $a,b$.

The calculations below simplify in the ``unitary gauge" $a=b=\infty$, thus in the following 
we will work in this ``gauge". In this case
\begin{equation}
j_\mu=\overline{u}^\alpha(p')(\Gamma_{\alpha\beta\nu\mu}p'^\nu
+\Gamma_{\alpha\beta\mu\nu}p^\nu)u^\beta(p)\equiv\overline{u}^\alpha(p')
{\cal O}(p',p)_{\alpha\beta\mu}u^\beta(p),
\end{equation}
and the electromagnetic vertex reads
\begin{eqnarray}\label{vertice}
{\cal V}(p',p,\infty)_{\alpha\beta\mu}&\equiv&{\cal O}(p',p)_{\alpha\beta\mu}=\Gamma_{\alpha\beta\nu\mu}p'^{\nu}+\Gamma_{\alpha\beta\mu\nu}p^{\nu},\\
\overline{\mathcal{O}}(p',p)_{\alpha\beta\mu}&\equiv&\gamma^0 [{\mathcal{O}}(p',p)_{\alpha\beta\mu}]^\dagger\gamma^0={\mathcal{O}}(p,p')_{\beta\alpha\mu}.
\end{eqnarray}
The propagator in this case is
\begin{equation}\label{propagador}
S(p,\infty,\infty)\equiv \Pi(p)=\frac{-\mathbf{P^{(\frac{3}{2})}}+\frac{p^2-m^2}{m^2}\mathbf{P}^{(\frac{1}{2})}}{p^2-m^2+i\epsilon}\equiv\frac{\Delta(p)}{p^2-m^2+i\epsilon},
\end{equation}
and the Ward-Takahashi identity simplifies to
\begin{equation}
(p'-p)^\mu \mathcal{O}(p',p)_{\alpha\beta\mu}
=\lbrace K_{\alpha\beta}(p')-m^2 g_{\alpha\beta}\rbrace-\lbrace K_{\alpha\beta}(p)-m^2 g_{\alpha\beta}\rbrace
=\Pi^{-1}_{\alpha\beta}(p')-\Pi^{-1}_{\alpha\beta}(p).
\end{equation}

\section{Compton Scattering}
\begin{figure}[ht]
\centering
\includegraphics[scale=0.25]{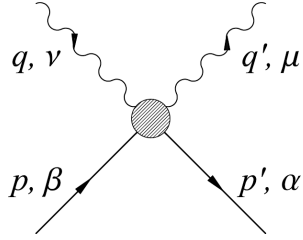}
\caption{Compton scattering off a spin 3/2 particle.}
\label{global}
\end{figure}

In this section we calculate Compton scattering. Our conventions are given in Fig (\ref{global}).
We will work in the rest frame of the initial spin 3/2 particle (lab frame). 
In this frame the differential cross section reads  
\aeq
\frac{\textup{d}\sigma}{\textup{d}\Omega}=\frac{1}{4(4\pi)^2}\frac{\vert
\overline{\M}\vert^2}{m^2}\left(\frac{\o'}{\o}\right)^2,
\label{comptondcs}
\ceq
where $m$ stands for the mass of the spin 3/2 particle and $\omega$, $\omega'$ denote the 
energies of the incoming and outgoing photon respectively. They are related by
\aeq
\o'=\frac{m\o}{m+\o(1-\cos\theta)},
\label{wp}
\ceq
where $\theta$ stands for the angle of the outgoing photon with respect to the incoming one. 

The amplitude for Compton scattering has three contributions: 
\begin{equation}\label{MNKR}
\mathcal{M}=\mathcal{M}_A+\mathcal{M}_B+\mathcal{M}_C
\end{equation}
where $\mathcal{M}_A,$ $\mathcal{M}_B$ and $\mathcal{M}_C$ correspond to $s$-channel, $u$-channel 
exchange and the ``seagull" contact term respectively: 
\begin{align}
\mathcal{M}_A=& e^2\overline{u}^{\alpha}(p^{\prime})\mathcal{O}%
(p^{\prime},Q)_{\alpha\gamma\mu}\Pi^{\gamma\delta}(Q)\mathcal{O}%
(Q,p)_{\delta\beta\nu}u^{\beta}(p)\epsilon^{\mu
*}(q^{\prime})\epsilon^{\nu}(q),\label{MANKR} \\
\mathcal{M}_B=&e^2\overline{u}^{\alpha}(p^{\prime})\mathcal{O}%
(p^{\prime},R)_{\alpha\gamma\nu}\Pi^{\gamma\delta}(R)\mathcal{O}%
(R,p)_{\delta\beta\mu}u^{\beta}(p)\epsilon^{\mu
*}(q^{\prime})\epsilon^{\nu}(q), \label{MBNKR}\\
\mathcal{M}_C=&-e^2\overline{u}^{\alpha}(p^{\prime})(\Gamma_{\alpha\beta\mu%
\nu}+\Gamma_{\alpha\beta\nu\mu})u^{\beta}(p)\epsilon^{\mu
*}(q^{\prime})\epsilon^{\nu}(q),\label{MCNKR}
\end{align}
with $Q=p+q=p^{\prime}+q^{\prime}$ and $R=p^{\prime}-q=p-q^{\prime}$. As a check, 
replacing $\epsilon^{\nu}(q) $ by
$q^{\nu}$ and using the Ward-Takahashi identity we obtain%
\begin{align}
{\cal M}_A(\epsilon^{\nu}(q)\rightarrow q^{\nu}) =&\quad e^2\overline{u}^{\alpha}(p'){\cal O}(p',Q)_{\alpha\beta\mu}u^{\beta}(p)\epsilon^{*\mu}(q'), \\
{\cal M}_B(\epsilon^{\nu}(q)\rightarrow q^{\nu}) =&-e^2\overline{u}^{\alpha}(p'){\cal O}(R,p)_{\alpha\beta\mu}u^{\beta}(p)\epsilon^{*\mu}(q'),\\
{\cal M}_C(\epsilon^{\nu}(q)\rightarrow q^{\nu})=&-e^2\overline{u}^{\alpha}(p')[{\cal O}(p',Q)_{\alpha\beta\mu}-
{\cal O}(R,p)_{\alpha\beta\mu}]u^{\beta}(p)\epsilon^{*\mu}(q').
\end{align}
Adding up these contributions we obtain that gauge invariance is satisfied \cite{Siahaan}
\begin{equation}
\mathcal{M}\left(  \epsilon(q)\rightarrow q\right)  =0.
\end{equation}
A similar result is obtained for the outgoing photon. 

The calculation of the spin averaged squared amplitude is straightforward but involve a 
large number of manipulations and properties of the formalism hence we will give some details. From Eqs. (\ref{MNKR},\ref{MANKR},\ref{MBNKR},\ref{MCNKR}) we obtain 
\begin{align}
\vert \overline{{\cal M}}\vert^2&=\frac{1}{8}\sum_{{pol}}\vert {\cal M}\vert^2\\
&=\frac{e^4}{8}Tr[
\tilde{\Delta}^{\eta\alpha}(p')\lbrace
{\cal O}(p',Q)_{\alpha\gamma\mu}\Pi^{\gamma\delta}(Q){\cal O}(Q,p)_{\delta\beta\nu}
+{\cal O}(p',R)_{\alpha\gamma\nu}\Pi^{\gamma\delta}(R){\cal O}(R,p)_{\delta\beta\mu}
-(\Gamma_{\alpha\beta\mu\nu}+\Gamma_{\alpha\beta\nu\mu})\rbrace\nonumber\\
&\quad\times\tilde{\Delta}^{\beta\zeta}(p)\lbrace
{\cal O}(p,Q)_{\zeta\phi\rho}\Pi^{\phi\theta}(Q){\cal O}(Q,p')_{\theta\eta\sigma}
+{\cal O}(p,R)_{\zeta\phi\sigma}\Pi^{\phi\theta}(R){\cal O}(R,p')_{\theta\eta\rho}
-(\Gamma_{\zeta\eta\rho\sigma}+\Gamma_{\zeta\eta\sigma\rho})\rbrace] g^{\mu\sigma} g^{\nu\rho}\nonumber.
\end{align}
Here, $\tilde{\Delta}$ denotes the projector onto the subspaces spanned by the desired 
solutions to the free equation 
\begin{equation}
\tilde{\Delta}_{\alpha\beta}(p)=\sum_{\lambda}u_{\alpha}(p,\lambda)\overline{%
u}_{\beta}(p,\lambda).
\end{equation}
Since we are working with parity-conserving interactions we will use the solutions with well 
defined parity. These solutions were constructed in \cite{NKR} and we just quote the 
final result here. 
\begin{align}
u^\alpha(\mathbf{p},3/2)=&\eta^\alpha(\mathbf{p},1)u(\mathbf{p},1/2),\qquad
u^\alpha(\mathbf{p},1/2)=\frac{1}{\sqrt{3}}\eta^\alpha(\mathbf{p},1)u(%
\mathbf{p},-1/2)+\sqrt{\frac{2}{3}}\eta^\alpha(\mathbf{p},0)u(\mathbf{p},1/2),
\\
u^\alpha(\mathbf{p},-1/2)=&\frac{1}{\sqrt{3}}\eta^\alpha(\mathbf{p},-1)u(%
\mathbf{p},1/2)+\sqrt{\frac{2}{3}} \eta^\alpha(\mathbf{p},0)u(\mathbf{p}%
,-1/2), \qquad u^\alpha(\mathbf{p},-3/2)=\eta^\alpha(\mathbf{p},-1)u(\mathbf{%
p},-1/2),
\end{align}
where 
\begin{align}
&\eta(\mathbf{p},1):=\frac{1}{\sqrt{2}m(m+p_0)} \left( 
\begin{array}{c}
-\left(m+p_0\right) \left(p_1+i p_2\right) \\ 
-m^2-p_0 m-p_1^2-i p_1 p_2 \\ 
-i \left(p_2^2-i p_1 p_2+m \left(m+p_0\right)\right) \\ 
-\left(p_1+i p_2\right) p_3%
\end{array}
\right),\quad \eta(\mathbf{p},0):=\frac{1}{m(m+p_0)} \left( 
\begin{array}{c}
\left(m+p_0\right) p_3 \\ 
p_1 p_3 \\ 
p_2 p_3 \\ 
p_3^2+m \left(m+p_0\right)%
\end{array}%
\right),  \notag \\
&\eta(\mathbf{p},-1):=\frac{1}{\sqrt{2}m(m+p_0)} \left( 
\begin{array}{c}
\left(m+p_0\right) \left(p_1-i p_2\right) \\ 
m^2+p_0 m+p_1^2-i p_1 p_2 \\ 
-i \left(p_2^2+i p_1 p_2+m \left(m+p_0\right)\right) \\ 
\left(p_1-i p_2\right) p_3%
\end{array}
\right),
\end{align}
and 
\begin{equation}
u\left(\mathbf{p},\frac{1}{2}\right):= \frac{1}{\sqrt{2m(m+p_0)}} \left( 
\begin{array}{c}
m+p_0 \\ 
0 \\ 
p_3 \\ 
p_1+i p_2%
\end{array}
\right),\quad u\left(\mathbf{p},-\frac{1}{2}\right):= \frac{1}{\sqrt{2m(m+p_0)}}%
\left( 
\begin{array}{c}
0 \\ 
m+p_0 \\ 
p_1-i p_2 \\ 
-p_3%
\end{array}
\right).
\end{equation}
Using these solutions, a straightforward calculation yields 
\begin{equation}
\tilde{\Delta}_{\alpha\beta}(p)=\sum_{\lambda}u_{\alpha}(p,\lambda)\overline{%
u}_{\beta}(p,\lambda)=-\Delta_{\alpha\beta}(p)\frac{\not{p}+m}{2m},
\label{Dtilde}
\end{equation}
where $\Delta_{\alpha\beta}(p)$ is the operator associated with the NKR
propagator in Eq.(\ref{propagador}). 

It is important to remark that the formalism we are using is based on the projection 
onto subspaces of the Casimir operators of the Poincar\'{e} group, $W^{2}$ and $P^{2}$. 
This projection does not define the parity properties of the solutions in the case of 
spin 3/2 (it does in the case of spin 1!) . However, it is always possible to choose 
solutions with well defined parity as we have done. In this case the external product 
of the solutions projects also onto the parity subspaces. This is the reason of the  
$(\not{p}+m)/2m$ factor in Eq. (\ref{Dtilde}). As a check we also constructed the 
negative parity solutions obtaining a similar result as  Eq. (\ref{Dtilde}) but with 
the factor $(-\not{p}+m)/2m$. 

In order to simplify the trace calculation by symmetry considerations, we use the notation 
\begin{equation}
\vert \mathcal{\overline{M}\vert}^2=AA+AB-AC+BA+BB-BC-CA-CB+CC,
\end{equation}
where 
\begin{align}
AA=&Tr[ \tilde{\Delta}^{\eta\alpha}(p^{\prime}) \mathcal{O}%
(p^{\prime},Q)_{\alpha\gamma\mu}\Pi^{\gamma\delta}(Q)\mathcal{O}%
(Q,p)_{\delta\beta\nu} \tilde{\Delta}^{\beta\zeta}(p) \mathcal{O}%
(p,Q)_{\zeta\phi\rho}\Pi^{\phi\theta}(Q)\mathcal{O}(Q,p^{\prime})_{\theta%
\eta\sigma} ] e^4 g^{\mu\sigma} g^{\nu\rho}/8, \\
AB=&Tr[ \tilde{\Delta}^{\eta\alpha}(p^{\prime}) \mathcal{O}%
(p^{\prime},Q)_{\alpha\gamma\mu}\Pi^{\gamma\delta}(Q)\mathcal{O}%
(Q,p)_{\delta\beta\nu} \tilde{\Delta}^{\beta\zeta}(p)\mathcal{O}%
(p,R)_{\zeta\phi\sigma}\Pi^{\phi\theta}(R)\mathcal{O}(R,p^{\prime})_{\theta%
\eta\rho} ] e^4 g^{\mu\sigma} g^{\nu\rho}/8, \\
AC=&Tr[ \tilde{\Delta}^{\eta\alpha}(p^{\prime}) \mathcal{O}%
(p^{\prime},Q)_{\alpha\gamma\mu}\Pi^{\gamma\delta}(Q)\mathcal{O}%
(Q,p)_{\delta\beta\nu} \tilde{\Delta}^{\beta\zeta}(p)(\Gamma_{\zeta\eta\rho%
\sigma}+\Gamma_{\zeta\eta\sigma\rho}) ] e^4 g^{\mu\sigma} g^{\nu\rho}/8, \\
CC=&Tr[ \tilde{\Delta}^{\eta\alpha}(p^{\prime})
(\Gamma_{\alpha\beta\mu\nu}+\Gamma_{\alpha\beta\nu\mu}) \tilde{\Delta}%
^{\beta\zeta}(p)(\Gamma_{\zeta\eta\rho\sigma}+\Gamma_{\zeta\eta\sigma\rho})
] e^4 g^{\mu\sigma} g^{\nu\rho}/8,
\end{align}
the other traces can be found using the following symmetry properties
\begin{equation}
AA\overset{u\leftrightarrow s}{=}BB,\qquad AB\overset{u\leftrightarrow s}{=}%
BA,\qquad AC\overset{u\leftrightarrow s}{=}BC, \qquad CA\overset{%
u\leftrightarrow s}{=}CB,
\end{equation}
so that we only need to calculate half the traces. We still have 
heavy calculation to carry out due the undetermined parameters 
in Eq. (\ref{gamageneral}). However, some of these
parameters must vanish if we want to preserve parity. Indeed, it can be shown 
that $\tilde{c}$ and $\tilde{d}$ yield odd-parity multipoles hence they must 
vanish in a parity invariant theory. 
With this simplification and using the constraints, the interaction current 
has a Gordon-like decomposition of the form 
\begin{equation}
j_\mu=\overline{u}^\alpha (p^{\prime})\left[g_{\alpha
\beta}(p^{\prime}+p)_\mu+ig [M_{\mu\nu}]_{\alpha\beta}(p^{%
\prime}-p)^\nu-i\gamma^5 f \epsilon_{\alpha\beta\mu\nu}(p^{\prime}-p)^\nu%
\right]u^\beta(p). 
\end{equation}
A final simplification consist in reducing all vertex functions appearing in
the trace by the projection rules 
\begin{equation}
\Delta^{\eta\alpha}(p)p_\alpha=\Delta^{\eta\alpha}(p)\gamma_\alpha=0,\qquad p_\eta\Delta^{\eta\alpha}(p)=\gamma_\eta\Delta^{\eta\alpha}(p)=0.
\end{equation}
After these simplifications we calculate the average squared amplitude with the aid 
of the FeynCalc package. The result is too long to be included here and we defer it 
to the appendix. It depends on the free parameters $f$ and $g$, on the
Mandelstam variables $s$ and $u$ and is manifestly crossing symmetric. In the lab frame
\begin{equation}
s=(p+q)^2=m(m+2\omega),\qquad t=(q^{\prime}-q)^2=-2\omega\omega^{\prime}(1-\cos \theta )
,\qquad u=(p-q^{\prime})^2=m(m-2\omega^{\prime}).
\end{equation}
The classical limit corresponds to the low energy limit $\omega\ll m$. The expansion of 
the average squared amplitude in this limit yields
\begin{equation}  
\frac{d\sigma(f,g,\eta,x)}{d\Omega}=
r_0^2 \left( \frac{x^2+1}{2}+\left(x^3-x^2+x-1\right) \eta +O\left(\eta ^2\right) \right),
\label{dsigclas}
\end{equation}
where $\eta=\omega/m$, $x=\cos\theta$ and $r_{0}=\alpha/m$ denotes the classical radius. 
Therefore, in the classical limit we obtain a differential cross section 
which is independent of the undetermined parameters and coincides with the Thomson result  
\begin{equation}
\left[\frac{d\sigma(f,g,\eta,x)}{d\Omega}\right]%
_{\eta\rightarrow 0}= \frac{1}{2}(1+x^2)r_0^2.
\end{equation}

\section{Compton scattering off Rarita-Schwinger spin $\frac{3}{2}$ particles.}

The Rarita-Schwinger Lagrangian is
\begin{eqnarray}
\mathcal{L}^{(RS)}(A) & =&\overline{\psi}^{\mu}
\left(  i\partial_{\alpha}\Gamma_{\mu~~\nu}^{~~\alpha}(A)- m~B_{\mu
\nu}(A)\right) \psi^{\nu}\, ,
\end{eqnarray}
where
\begin{eqnarray}
\Gamma_{\mu~~\nu}^{~~\alpha}(A)&=&g_{\mu\nu}\gamma_{\alpha}
+A(\gamma_{\mu}g^{\alpha}_{~\nu}+g_{\mu}^{~\alpha}\gamma_{\nu})
+B\gamma_{\mu} \gamma^{\alpha}\gamma_{\nu},  \nonumber \\
B_{\mu\nu}(A)&=&g_{\mu\nu}-C\gamma_{\mu}\gamma_{\nu} ,
 \label{Teil_2}\\ 
A\neq\-\frac{1}{2}, \quad B&\equiv&\frac{3}{2}A^{2}+A+\frac{1}{2}, \quad
C\equiv3A^{2}+3A+1. \nonumber %
\end{eqnarray}
The case $A=-\frac{1}{3}$ corresponds to the Lagrangian 
originally proposed in \cite{RS}, while for $A=-1$ the Lagrangian simplifies to 
\begin{equation}
\mathcal{L}^{(RS)}(A=-1) =\overline{\psi}^{\mu}
\left(  i\partial_{\alpha}\epsilon_{\mu~~\nu\rho}^{~~\alpha}
\gamma^{5}\gamma^{\rho}
- im~\sigma_{\mu\nu}\right) \psi^{\nu}. \label{super}
\end{equation}
The propagator is
\begin{equation}
\Delta_{\mu\nu} (p , A)= \frac{\Sigma_{\mu\nu} (p,A)}{p^2-m^2+ i \epsilon},
\end{equation}
with
\begin{equation}
\Sigma_{\mu\nu} (p,A)=2m S_{\mu\nu} - \frac{1}{6} ~\frac{A+1}{2A+1} 
~\frac{p^2- m^2}{m} \left\{ \gamma_\mu \bigg(\frac{2p}{m}-\gamma \bigg)_\nu 
 + \bigg(\frac{2p}{m}- \gamma \bigg)_\nu \gamma_\mu 
- \frac{A+1}{2A+1} \bigg( \gamma_\mu \frac{\not p}{m}\gamma_\nu -2\gamma_\mu
\gamma_\nu \bigg) \right\} 
\end{equation}
where
\begin{equation}
S_{\mu\nu}= \left\{- g_{\mu\nu}+ \frac{1}{3} \gamma_\mu \gamma_\nu - \frac{1}{3m}
(\gamma_\mu p_\nu- p_\mu \gamma_\nu)+\frac{2}{3m^2} p_\mu p_\nu \right\} 
\frac{\not p +m}{2m}.
\end{equation}
Electromagnetic interactions are introduced using the 
gauge principle which amounts to use the minimal coupling 
$\partial_{\mu}\to D_{\mu}= \partial_{\mu}-ieA_{\mu}$. The interacting Lagrangian is
\begin{equation}
{\cal L}_{int}=\overline{\psi}^{\alpha }\left[i D^{\mu }\Gamma _{\alpha \mu \beta }(A)
-m B_{\alpha \beta }(A)\right]\psi ^{\beta }.
\end{equation}
The electromagnetic current reads
\begin{equation}
j_{\mu}=\overline{\psi}^{\alpha}\Gamma_{\alpha\mu\beta}(A)\psi^{\beta},%
\end{equation}
which yields the vertex function
\begin{equation}
\mathcal{O}_{\alpha\beta\mu}(A)=\Gamma_{\alpha\mu\beta}(A).
\end{equation}
If we define 
\begin{equation}
{\cal K}_{\mu\nu}(p,A)= p_{\alpha}\Gamma_{\mu~~\nu}^{~~\alpha}(A)- 
m~B_{\mu\nu}(A),
\end{equation}
it can be easily shown that the Ward-Takahashi identity holds
\begin{equation}
(p^{\prime}-p)^{\mu}\mathcal{O}_{\alpha\beta\mu}(A)={\cal K}_{\alpha\beta}(p^{\prime
},A)-{\cal K}_{\alpha\beta}(p,A).
\end{equation}
The interacting Lagrangian can be factorized as
\begin{equation}
\mathcal{L}^{(RS)}(A)  =\overline{\psi}^{\mu}~
R_{\mu\rho}\left(\frac{A}{2}\right)~
{\cal K}^{\rho\sigma}(\pi,0)R_{\sigma\nu}\left(\frac{A}{2}\right) \psi^{\nu}\, ,
\end{equation}
where $\pi_{\mu}=p_{\mu}-eA_{\mu}$ and 
\begin{equation}
R_{\mu\rho}(w)\equiv g_{\mu\rho}+w \gamma_{\mu}\gamma_{\rho}.
\label{rotation}
\end{equation}
This factorization can be used to show that the Lagrangian is 
invariant under the point transformations 
\begin{equation}
\psi_{\mu}\to\psi^{\prime}_{\mu}=R_{\mu\nu}(w)\psi^{\nu},\quad 
A\to \frac{A-2w}{1+4w}. \label{point}
\end{equation}
The freedom represented by the parameter $A$ reflects
 invariance under ``rotations" mixing the two spin-$\frac{1}{2}^{+}$ and 
 $\frac{1}{2}^{-}$ sectors residing in the RS representation space besides 
spin-$\frac{3}{2}$ \cite{todos}. 
It can be shown  \cite{KOS} that the elements of the $S$ matrix 
do not depend on the parameter $A$. In the following we will work 
with $A=-1$ in whose case the propagator takes its simplest form.
 
Compton scattering is induced by the $s$ and $u$ channel conventional diagrams. The 
corresponding amplitudes are
\begin{eqnarray}
\mathfrak{M}_{s}&=&e^2\bar{u}^{\alpha }(p')\mathcal{O}_{\alpha \gamma \mu }\Pi ^{\gamma \delta }(Q)
\mathcal{O}_{\delta \beta \nu }u^{\beta }(p)\epsilon ^{\nu }(q) \epsilon^{*\mu }(q')\\
\mathfrak{M}_{u}&=&e^2\bar{u}^{\alpha }(p')\mathcal{O}_{\alpha \gamma \nu }\Pi ^{\gamma \delta }(R)
\mathcal{O}_{\delta \beta \mu }u^{\beta }(p)\epsilon ^{\nu }(q)\epsilon^{*\mu }(q').
\end{eqnarray}
Replacing $\epsilon^{\nu}(q) $ by $q^{\nu}$ and using the Ward-Takahashi identity we obtain%
\begin{eqnarray}\label{mab}\nonumber
\mathfrak{M}_{s}(\epsilon (q)\to q)
&=&e^2\epsilon ^{\mu }(q')\bar{u}^{\alpha }(p')\mathcal{O}_{\alpha \beta \mu }u^{\beta }(p),\\
\mathfrak{M}_{u}(\epsilon (q)\to q)
&=&-e^2\bar{u}^{\alpha }(p')\mathcal{O}_{\alpha \beta \mu }u^{\beta }(p)\epsilon ^{\mu }(q'),
\end{eqnarray}
and gauge invariance is obtained adding up Eqs.(\ref{mab}). Analogous results 
hold for the outgoing photon. 
 
The average squared amplitude is obtained using the FeynCalc package as
\begin{eqnarray}\label{RSasa}
|\bar{\mathfrak{M}}_{RS}|^{2}&=&\frac{ e^{4}}{81 m^8 \left(m^2-s\right)^2 \left(m^2-u\right)^2}
[1530 m^{16}-996 (s+u) m^{14}+\left(59 s^2-982 u s+59 u^2\right) m^{12}  \\ \nonumber
&+&12 (s+u) \left(8 s^2+63 u s+8 u^2\right) m^{10}-\left(48 s^4+269 u
   s^3+358 u^2 s^2+269 u^3 s+48 u^4\right) m^8 \\  \nonumber
&+&(s+u) \left(19 s^4+56 u s^3+142 u^2 s^2+56 u^3 s+19 u^4\right) m^6 \\ \nonumber
&-& s u \left(24 s^4+37 u s^3+94 u^2
   s^2+37 u^3 s+24 u^4\right) m^4  \\ \nonumber
   &+& 2 s^2 u^2 (s+u) \left(3 s^2+8 u s+3 u^2\right) m^2-s^3 u^3 \left(s^2+u^2\right)].
\end{eqnarray}
It is explicitly symmetric under $s\leftrightarrow u$ exchange. In the lab frame the differential 
cross section reads
\begin{eqnarray}\label{dsigmaRS}
\frac{d\sigma_{RS}}{d\Omega}&=&
\frac{r^{2}_{0}}{162 (1+ \eta (1-x) )^5}[2 (x-1)^2 \left(15 x^2-36 x+25\right) \eta ^6 \\ \nonumber
&-& 2 (x-1) \left(3 x^4-16 x^3+134 x^2-216 x+103\right) \eta ^5 
+ \left(7 x^4-244 x^3+1010 x^2-1284 x+527\right) \eta ^4 \\ \nonumber
&-& (x-1) \left(81 x^4-162 x^3+164 x^2-582 x+723\right) \eta ^3
+\left(243 x^4-486 x^3+487 x^2-696 x+564\right) \eta ^2 \\ \nonumber
&-& 243 (x-1)\left(x^2+1\right) \eta +81 \left(x^2+1\right)].  
\end{eqnarray}
In the low energy limit we get 
\begin{equation}  
\frac{d\sigma_{RS}}{d\Omega}=
r_0^2 \left( \frac{x^2+1}{2}+\left(x^3-x^2+x-1\right) \eta +O\left(\eta ^2\right) \right),
\label{dsigleRS}
\end{equation}
and comparing with Eq. (\ref{dsigclas}) we can see that the predictions of the RS 
and NKR formalisms coincide to order $\eta$. In particular, in the classical limit 
the Thomson result is obtained in both formalisms. Integrating the solid angle we 
get the total cross section
\begin{eqnarray} \label{csRS}
\sigma_{RS}&=&
\frac{\sigma_{T}}{648 \eta ^3 (2 \eta +1)^4}[3 \left(30 \eta ^4+8 \eta ^3-23 \eta ^2
-162 \eta -162\right) \log (2 \eta +1) (2 \eta +1)^4 \\ \nonumber
&+& 2 \eta  \left(144 \eta ^9+232 \eta ^8+1444 \eta
   ^7+4344 \eta ^6+8182 \eta ^5+15510 \eta ^4+18927 \eta ^3+12219 \eta ^2+3888 \eta +486\right)],  
\end{eqnarray}
where $\sigma_{T}=8\pi r^{2}_{0}/3$ stands for the Thomson total cross section. As far as we know, these results were obtained firstly in \cite{DT} using a different procedure.  

\section{Discussion}
Before we start the discussion of our results it is important to recall results for 
Compton scattering of particles with lower spin. In the case of scalar particles a 
straightforward calculation yields
\begin{eqnarray}\label{ASA0}
|\overline{\mathfrak{M}}_{0}|^{2}&=&\frac{4e^{4} \left(5 m^8-4 (s+u) m^6+\left(s^2+u^2\right) m^4+s^2
   u^2\right)}{\left(m^2-s\right)^2 \left(m^2-u\right)^2}.
\end{eqnarray}

For a Dirac particle we obtain
\begin{eqnarray}\label{ASA12}
|\overline{\mathfrak{M}}_{\frac{1}{2}}|^{2}&=& \frac{4 e^{4}\left(6 m^8-\left(3 s^2+14 u s+3 u^2\right)
   m^4+\left(s^3+7 u s^2+7 u^2 s+u^3\right) m^2-s u
   \left(s^2+u^2\right)\right)}{2\left(m^2-s\right)^2
   \left(m^2-u\right)^2}.
\end{eqnarray} 

The calculation of Compton scattering in the NKR formalism for a vector particle, i.e., 
a spin 1 particle transforming in the $(1/2,1/2)$ representation of the Homogeneous 
Lorentz Group, was done in \cite{NRDK}. The electromagnetic structure of a vector particle 
is characterized by two free parameters $g$ and $\xi$, the last one corresponding to the 
odd-parity terms. The specific values of $g$ and $\xi$ were fixed in \cite{NRDK} analyzing 
the high energy behavior of the total cross section for Compton scattering and it was concluded 
there that the only values preserving unitarity in the high energy limit are $g=2$ and $\xi=0$.
As discussed in \cite{NRDK} these values reproduce the electromagnetic couplings of 
the $W$ boson in the Standard Model. The average squared amplitude in this case turns out to be
\begin{eqnarray}\label{ASA1}
|\overline{\mathfrak{M}}_{1}|^{2}&=&\frac{4 e^{4}}{3 \left(m^2-s\right)^2
   \left(m^2-u\right)^2}[31 m^8-44 m^6 (s+u)+m^4 \left(31 s^2+40 s
   u+31 u^2\right)\\ \nonumber
   &-&4 m^2\left(3 s^3+5 s^2 u+5 s u^2+3 u^3\right)+2 s^4+4 s^3 u+7 s^2 u^2+4 s u^3+2 u^4].
\end{eqnarray} 

The average squared amplitudes for spin $s=0,1/2,1$ in Eqs. (\ref{ASA0},\ref{ASA12},\ref{ASA1}) 
are symmetric under $s \leftrightarrow u$ exchange and have the interesting property that 
in the forward direction ($t=0$, $u=2m^{2}-s$) they are energy-independent and have the common 
value
\begin{equation}
|\overline{\mathfrak{M}}_{s}|^{2}_{forward}= 4e^{4}.
\label{eind}
\end{equation}
As can be seen using Eq. (\ref{comptondcs}), in the rest frame of a particle with spin 
$s=0,1/2,1$, it requires the differential cross section for Compton scattering in the 
forward direction to be energy independent and coincide with the classical squared radius 
\begin{equation}
\left.\frac{d\sigma_{s}}{d\Omega}\right|_{forward}=r_0^2.
\label{condition}
\end{equation}

As for spin $3/2$, the Rarita-Schwinger result quoted in Eq.(\ref{RSasa}) 
in the forward direction reduces to
\begin{equation}\label{RSforward}
|\overline{\mathfrak{M}}_{RS}|^{2}_{forward}=\frac{2  e^{4}\left(191 m^8-60 m^6 s+34 m^4 s^2-4 m^2 s^3+s^4\right)}{81 m^8}.
\end{equation} 

In the NKR formalism the average squared amplitude in the appendix depends on two parameters 
$f$ and $g$ which determine the electromagnetic structure at tree level of the spin $3/2$ 
particle. It was shown in \cite{NKR} that causal propagation of spin $3/2$ waves in an electromagnetic background is obtained for $g=2$ and $f=0$ and we will consider these values in the following.
Using these values we get the average squared amplitude as
\begin{eqnarray}\label{ASANKR}\nonumber
|\overline{\mathfrak{M}}_{NKR}|^{2}&=&\frac{4 e^{4}}{81 m^{10}\left(m^2-s\right)^2 \left(m^2-u\right)^2}
   [5952 m^{18}-5272 m^{16} (s+u)+m^{14} \left(310 s^2-6148 s u+310 u^2\right)\\ \nonumber
   &+& 2 m^{12} \left(1045 s^3+5703 s^2 u+5703 s u^2+1045 u^3\right) \\ \nonumber
   &-& m^{10} \left(1401 s^4+7048 s^3 u+9966 s^2 u^2+7048 s u^3+1401 u^4\right)\\ 
   &+& m^8 \left(339 s^5+2119 s^4 u+3718 s^3 u^2+3718 s^2 u^3+2119 s u^4+339 u^5\right)\\ \nonumber
   &-&m^6 \left(22 s^6+343 s^5 u+764 s^4 u^2+678 s^3 u^3+764 s^2 u^4+343 s u^5+22 u^6\right)\\ \nonumber
   &+& 2 m^4 \left(s^7+26 s^6 u+93 s^5 u^2+52 s^4 u^3+52 s^3 u^4+93 s^2 u^5+26 s u^6+u^7\right)\\ \nonumber
   &-& 4 m^2 s u \left(s^6+8 s^5 u+6 s^4 u^2+2 s^3 u^3+6 s^2 u^4+8 s u^5+u^6\right) 
   + 2 s^2 u^2 (s+u) \left(s^2+u^2\right)^2].
\end{eqnarray} 
In the forward direction this average squared amplitude has the value 
\begin{equation}
|\overline{{\mathfrak{M}}}_{NKR}|^{2}= 4e^{4}.
\end{equation}
We remark that the properties in Eqs. (\ref{eind},\ref{condition}) 
are satisfied by a spin $3/2$ in the NKR formalism but not in the RS formalism.

\begin{figure}[tbp]
\begin{center}
\includegraphics{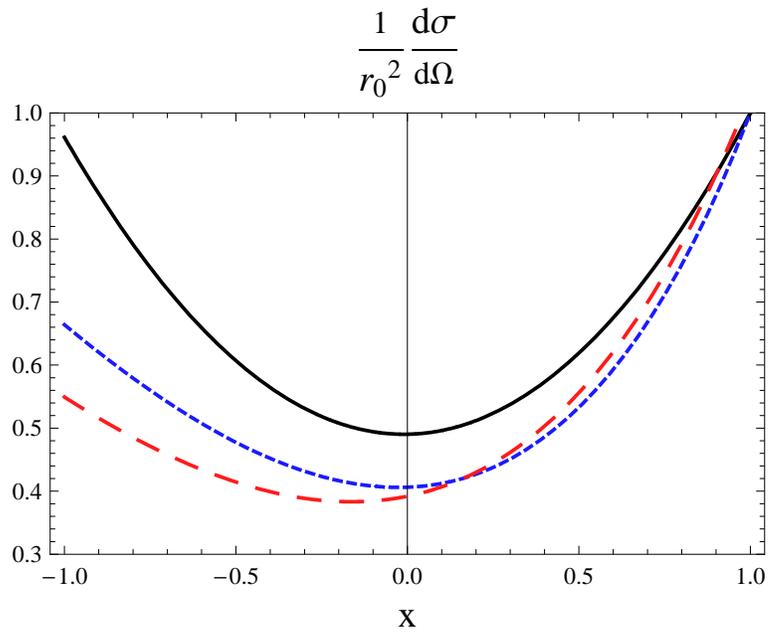}
\end{center}
\caption{Differential cross section in the RS and NKR formalisms  as a function of 
$x=\cos\protect\theta$ for low values of the
energy of the incident photon in the laboratory frame: $\protect\eta=\omega/m$. 
The black curve corresponds to $\protect\eta=0$ (Thomson limit), dashed curves 
correspond to $\protect\eta=0.2$ in the NKR formalism (sort-dashed curve) and 
RS formalism (long-dashed curve).}
\label{dsigma32f0g2}
\end{figure}

\begin{figure}[tbp]
\begin{center}
\includegraphics{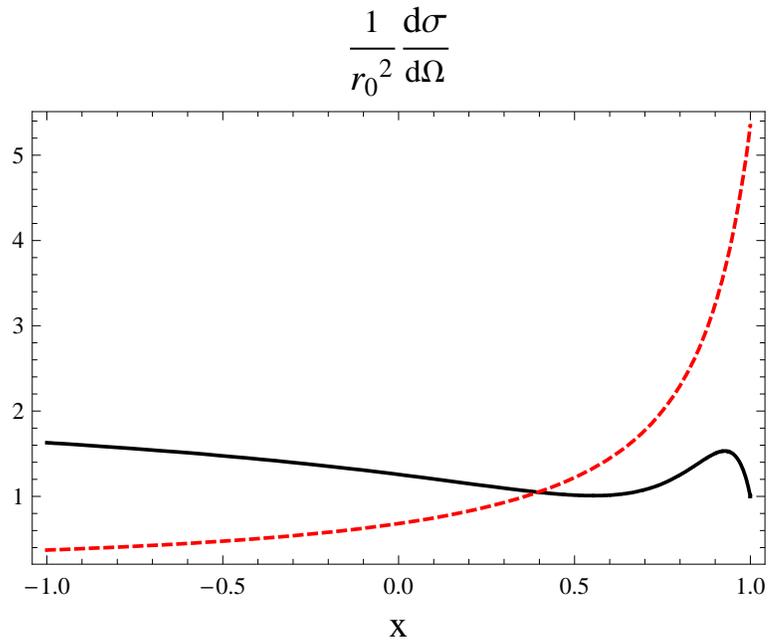}
\end{center}
\caption{Differential cross section in the RS and NKR formalisms  as a function of 
$x=\cos\protect\theta$ for $\protect\eta=1.5$. The solid curve corresponds to the 
results of the NKR formalism while the dashed curve are the results of the RS formalism.}
\label{high}
\end{figure}

The differential cross section reads
\begin{equation}
\frac{d\sigma_{NKR}}{d\Omega}=\frac{r_0^2}{(1+\eta z)^{7}}\sum_{n=0}^{10} h_{n}(z)\eta^{n}.
\label{dcsNKR}
\end{equation}
with $z=1-x$, $x=\cos \theta$ and
\begin{eqnarray}
h_{0}(z)&=&\frac{1}{2} \left(z^2-2 z+2\right),  \\ \nonumber
h_{1}(z)&=& \frac{5}{2} z \left(z^2-2 z+2\right), \\ \nonumber
h_{2}(z)&=& \frac{5}{2} z^2 \left(2 z^2-4 z+5\right), \\ \nonumber
h_{3}(z)&=& 5 z^3 \left(z^2-2 z+4\right), \\ \nonumber
h_{4}(z)&=&\frac{1}{18} z^2 \left(45 z^4-90 z^3+384 z^2+8 z+18\right), \\ \nonumber
h_{5}(z)&=& \frac{1}{6} z^3 \left(3 z^4-6 z^3+90 z^2+8 z+18\right), \\ \nonumber
h_{6}(z)&=& \frac{1}{162} z \left(1053 z^5+297 z^4+549 z^3+108 z^2-208 z+64\right), \\ \nonumber
h_{7}(z)&=& \frac{1}{81} z^2 \left(108 z^5+117 z^4+144 z^3+108 z^2-208 z+64\right),\\ \nonumber
h_{8}(z)&=& \frac{1}{162} z^3 \left(81 z^4+63 z^3+120 z^2-256 z+128\right), \\ \nonumber
h_{9}(z)&=& \frac{2}{81} z^4 \left(3 z^2-12 z+16\right), \\ \nonumber
h_{10}(z)&=& \frac{8 z^5}{81}.  \\ \nonumber
\label{hs}
\end{eqnarray} 
\begin{figure}[tbp]
\begin{center}
\includegraphics{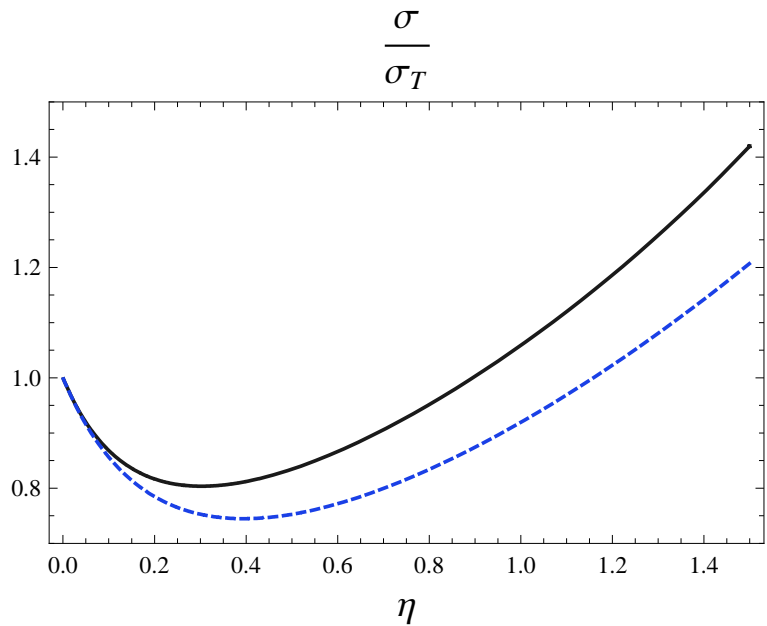}
\end{center}
\caption{Total cross section normalized to the Thomson cross section. The solid line 
corresponds to the NKR formalism. The dashed curve is the result of the RS formalism.}
\label{sigma32f0g2}
\end{figure}
In Fig. \ref{dsigma32f0g2} we show the results of both formalisms for the differential 
cross section for low values of $\eta$. Although both formalisms coincide in the classical limit, 
even for values as low as $\eta=0.2$ there are sizable differences in the angular distribution 
of the emitted photons. For higher values of $\eta$ these differences become more important 
as shown in Fig. (\ref{high}).

Integrating the solid angle we find the total cross section as 
\begin{eqnarray}\label{sigtot}
\sigma_{NKR}&=&\frac{\sigma_{T}}{3240\eta ^{3}(2\eta +1)^{6}}[45(2\eta +1)^{6}\left( 4\eta
^{5}+21\eta ^{4}-111\eta ^{3}-153\eta ^{2}-54\eta -54\right) \log (2\eta +1) \nonumber \\
&+&2\eta 
( 
5376\eta ^{12}-640\eta ^{11}-15936\eta ^{10}+14984\eta ^{9}+516640\eta
^{8}+1467750\eta ^{7} + 2010150\eta ^{6} \nonumber \\ 
&+& 1742445\eta ^{5}+1082160\eta ^{4}+493830\eta
^{3}+155115\eta ^{2}+29160\eta +2430
) ].
\end{eqnarray}
The cross section normalized to the Thomson one is shown in Fig. (\ref{sigma32f0g2}) for 
$\eta\leq 1.5$ along with the result of the RS formalism in Eq. (\ref{csRS}). The NKR and RS 
formalisms yield the same result in the Thomson limit but their predictions for the total 
cross section differ beyond this point.

In the high energy limit the total cross section predicted by the NKR formalism grows as 
$\eta^{4}$. This is in contrast with the spin 1 case studied in \cite{NRDK} where the total 
cross section remains finite in the high energy limit and further work is necessary in 
order to understand this point.

\section{Summary and perspectives}
In this work we study Compton scattering off a spin 3/2 elementary target in a 
recently proposed formalism for the description of high spin fields based on the 
Poincar\'{e} projectors and also in the conventional Rarita-Schwinger formalism. 
These formalisms yield the same result for the angular distribution and total cross 
section in the classical limit and coincide with the Thomson result. However, 
we obtain different predictions for these observables beyond this point, these 
differences becoming stronger at higher energies. 

It is pointed out that the average squared amplitudes for Compton scattering 
in the forward direction for lower spin, ($s=0,1/2,1$), are energy independent 
and have the common value $4 e^{4}$. In consequence, the differential cross sections 
in the forward direction and in the rest frame of the particles, coincide with the 
squared classical radius . This property is shared by the average squared 
amplitude for Compton scattering off spin $3/2$ particle as calculated 
in the Poincar\'{e} projector formalism but not in the Rarita-Schwinger formalism.

The classical regime tests only the lowest multipole (the electric charge), thus 
the differences in the angular distributions in these formalisms arise from the 
different predictions of these theories for higher multipoles and 
a calculation of these multipoles is desirable. Such analysis could also shed light 
on the high energy behavior of the total cross section. In contrast 
to the case of spin $1$ in the (1/2,1/2) representation space studied in \cite{NRDK}
which reproduces the electromagnetic couplings of the $W$ in the Standard Model and 
whose total cross section for Compton scattering remains finite at high energies, 
in the case of spin $3/2$ studied here it grows as $(\frac{\o}{m})^{4}$ in this energy regime.  

On the other hand, in the case of spin $1$  the Poincar\'{e} projectors automatically 
project onto subspaces with well defined parity. This is not the case for spin $3/2$ 
in whose case solutions with well defined parity must be chosen by hand. Therefore, 
it would be interesting to explore the consequences of a simultaneous projection 
onto well defined parity subspaces at the free particle level.
Under $U(1)_{em}$ gauging we expect different predictions for the higher multipoles 
in this case. 

\begin{acknowledgments}
Work supported by CONACyT-M\'{e}xico under project CONACyT-50471-F and DINPO-UG. We thank 
Mariana Kirchbach and Sim\'{o}n Rodr\'{i}guez for useful conversations on this topic.
\end{acknowledgments}

\section{Appendix}
Our calculation yields the average squared amplitude
\begin{equation}
|\overline{\mathcal{M}}|^{2}=\frac{1}{324m^{10}\left( m^{2}-s\right)
^{2}\left( m^{2}-u\right) ^{2}}\sum_{n=0}^9{l_{2n}m^{2n}},
\end{equation}
where
{\small
\begin{align}
l_0=&2\left( f^{2}+gf+g^{2}\right) ^{2}s^{2}u^{2}(s+u)\left(
s^{2}+u^{2}\right) ^{2},\nonumber\\
l_2=&2 \left(-5 f^4+4 (9-10 g) f^3+4 \left(-9 g^2+6 g+7\right) f^2-4 g (g (g+12)-16) f+4 g^2 ((g-12) g+16)\right) s^4 u^4\nonumber\\
&-\left(25 f^4+(92 g-52) f^3+6 (g
   (25 g-16)-2) f^2+8 g (g (19 g-24)+6) f+4 g^2 (g (19 g-32)+12)\right)\nonumber\\
   &\times s^3 \left(s^2+u^2\right) u^3-4 \left(f^2+g f+g^2\right) \left(6 f^2+6 (3 g-2)
   f+3 g (5 g-4)-4\right) s^2 \left(s^4+u^4\right) u^2\nonumber\\
   &-4 \left(f^2+g f+g^2\right)^2 s \left(s^6+u^6\right) u\nonumber\\
l_4=&(s+u) \left(-2 \left(17 f^4+4 (38-17 g) f^3+\left(-303 g^2+852 g-504\right) f^2+4 (g (7 (27-5 g) g-276)+80) f\right.\right.\nonumber\\
&\left.\left.-4 (g (g (g (7
   g-136)+357)-316)+116)\right) s^3 u^3+\left(95 f^4+14 (35 g-32) f^3+(3 g (441 g-596)+448) f^2\right.\right.\nonumber\\
&\left.\left.+4 (g (g (316 g-459)+94)+24) f+2 (g (g (g (277
   g-592)+518)-504)+328)\right) \left(u^2 s^4+u^4 s^2\right)\right.\nonumber\\
&\left.+2 \left(f^2+g f+g^2\right) \left(17 f^2-36 f+44 g^2+8 (7 f-3) g-28\right) \left(u
   s^5+u^5 s\right)+2 \left(f^2+g f+g^2\right)^2 \left(s^6+u^6\right)\right)\nonumber\\
l_6=&-2 \left(185 f^4+2 (899 g-1126) f^3+(69 g (83 g-156)+4996) f^2+4 (g (g (1210 g-2709)+2272)-388) f\right.\nonumber\\
&\left.+g (g (g (2291 g-9592)+15856)-10448)+2976\right) s^3
   u^3-\left(447 f^4+30 (93 g-106) f^3+2 \left(4443 g^2\right.\right.\nonumber\\
&\left.\left.-7332 g+2846\right) f^2+4 (g (3 g (653 g-1150)+2062)-352) f+16 (g (g (3 g (76
   g-231)+892)-554)\right.\nonumber\\
&\left. +200)\right) \left(u^2 s^4+u^4 s^2\right)-\left(206 f^4+(826 g-736) f^3+(3 g (739 g-844)+584) f^2+4 (g (g (460 g-387)\right.\nonumber\\
&\left.-16)+60) f+g
   (g (g (941 g-1592)+968)-912)+1120\right) \left(u s^5+u^5 s\right)\nonumber\\
&-2 \left(f^2+g f+g^2\right) \left(16 g^2+22 f g+f (7 f-12)-20\right)
   \left(s^6+u^6\right)\nonumber\\
l_8=&(s+u) \left(2 \left(388 f^4+(3446 g-4020) f^3+2 \left(5079 g^2-9216 g+4054\right) f^2+8 \left(1165 g^3-2610 g^2+1792 g\right.\right.\right.\nonumber\\
&\left.\left.\left.-190\right) f+4849 g^4-15312
   g^3+18704 g^2-7984 g+1568\right) s^2 u^2+2 \left(349 f^4+2 (679 g-836) f^3\right.\right.\nonumber\\
&\left.\left.+\left(5175 g^2-7392 g+3232\right) f^2+4 \left(1013 g^3-1320 g^2+532
   g+80\right) f+2278 g^4-4628 g^3+4492 g^2\right.\right.\nonumber\\
&\left.\left.-1856 g+560\right) \left(u s^3+u^3 s\right)+\left(116 f^4+(202 g-268) f^3+6 \left(151 g^2-84 g+6\right)
   f^2\right.\right.\nonumber\\
&\left.\left.+8 \left(56 g^3-27 g^2-6 g+18\right) f+389 g^4-272 g^3+540 g^2-768 g+752\right) \left(s^4+u^4\right)\right)\nonumber\\
l_{10}=&-2 \left(1177 f^4+22 (359 g-386) f^3+(3 g (9953 g-14396)+15572) f^2+4 (g (g (6491 g-11139)+3668)\right.\nonumber\\
&\left.+2252) f+g (g (g (13105
   g-20512)+5072)+9424)-4992\right) s^2 u^2-\left(1647 f^4+36 (187 g-223) f^3\right.\nonumber\\
&\left.+2 (27 g (617 g-768)+7726) f^2+8 \left(g \left(3177 g^2-4362
   g+994\right)+1376\right) f+2 \left(g \left(g \left(6705 g^2-8688 g\right.\right.\right.\right.\nonumber\\
&\left.\left.\left.\left.+3604\right)+2800\right)-1408\right)\right) \left(u s^3+u^3 s\right)+\left(-416
   f^4-26 (5 g-28) f^3-(9 g (515 g-268)+160) f^2\right.\nonumber\\
&\left.-4 (g (g (463 g-249)+316)+76) f-g (g (g (1517 g-976)+2080)-1552)-736\right) \left(s^4+u^4\right)\nonumber\\
l_{12}=&\left(255 f^4+2 (2409 g-928) f^3+\left(55731 g^2-48972 g+5904\right) f^2+4 \left(10989 g^3-9339 g^2-5418 g+8632\right) f\right.\nonumber\\
&\left.+2 \left(6489 g^4+5212
   g^3-24534 g^2+21064 g+1736\right)\right) \left(u s^2+u^2 s\right)+\left(301 f^4-2 (653 g-544) f^3\right.\nonumber\\
&\left.+9 \left(1425 g^2-388 g-336\right) f^2+4
   \left(1511 g^3+231 g^2+258 g-72\right) f\right.\nonumber\\
&\left.+1414 g^4+2632 g^3-4404 g^2+4080 g-784\right) \left(s^3+u^3\right)\nonumber\\
l_{14}=&2 \left(1699 f^4+4 (908 g-2063) f^3-84 \left(276 g^2-44 g-75\right) f^2-16 \left(1108 g^3+603 g^2-1536 g+450\right) f\right.\nonumber\\
&\left.+2 \left(1649 g^4-11240
   g^3+14328 g^2-5376 g-7616\right)\right) s u+\left(1361 f^4+(5488 g-8756) f^3-4 \left(4854 g^2\right.\right.\nonumber\\
&\left.\left.+1056 g-2321\right) f^2-16 \left(704 g^3+1035 g^2-473
   g-608\right) f\right.\nonumber\\
   &\left.+2 \left(2407 g^4-8072 g^3+8584 g^2-2848 g-96\right)\right) \left(s^2+u^2\right)\nonumber\\
l_{16}=&-4 \left(735 f^4+(2178 g-3704) f^3+(1688-33 g (111 g+124)) f^2+4 (g (566-3 g (197 g+735))+2020) f\right.\nonumber\\
&\left.+g (g (g (3033 g-5308)+926)+4184)+2952\right) (s+u)\nonumber\\
l_{18}=&12 \left(139 f^4+(398 g-668) f^3-3 \left(117 g^2+316 g+20\right) f^2-4 \left(55 g^3+450 g^2+54 g-628\right) f\right.\nonumber\\
&\left.+4 \left(157 g^4-154 g^3-189 g^2+404
   g+652\right)\right)
\end{align}
}
This amplitude is clearly symmetric under the $s \leftrightarrow u$ exchange.

\end{document}